\documentclass[twoside,reqno]{mbook}
\usepackage{epsfig,cite}
\usepackage{amssymb,amsmath}
\usepackage{times}
\begin{document}

\sloppy \raggedbottom

\setcounter{page}{1}

\def\bra{\langle }
\def\ket{\rangle }
\def\bq{\begin{eqnarray}}
\def\eq{\end{eqnarray}}
\def\be{\begin{eqnarray}}
\def\ee{\end{eqnarray}}

\title{$^3$He structure and Generalized Parton
Distributions \thanks{Support from the funds
COFIN03 of MIUR is gratefully acknowledged.}}

\runningheads{$^3$He structure and Generalized Parton
Distributions}
{S.~Scopetta}

\begin{start}
\author{S. Scopetta}{1}

\address{Dipartimento di Fisica, Universit\`a degli Studi
di Perugia, via A. Pascoli
06100 Perugia, Italy
and INFN, sezione di Perugia}{1}

\begin{Abstract}

Nuclear
Generalized Parton Distributions (GPDs), a unique
tool to access several crucial features of nuclear structure,
could be measured in the coherent channel of
hard exclusive processes, such as deep electroproduction
of photons and mesons off nuclear targets.
Here, a realistic microscopic calculation
of the unpolarized quark GPD $H_q^3$
of the $^3$He nucleus is described.
In Impulse Approximation, 
$H_q^3$ is shown to be given by
a convolution between the GPD of the internal nucleon and
the non-diagonal spectral function,
describing properly
Fermi motion 
and binding effects.
The obtained formula has the correct
limits.
Nuclear effects, evaluated by a
modern realistic potential, 
are found to be larger than in the forward case.
In particular, they
increase with increasing the momentum
transfer and the asymmetry of the process.
Besides, it is found that the 
nuclear GPD cannot be factorized into a $\Delta^2$-dependent
and a $\Delta^2$-independent term, as suggested
in prescriptions proposed for finite nuclei.
The dependence of the obtained GPDs on different realistic potentials
used in the calculation shows that these quantities are
sensitive to the details of nuclear structure at short distances.
\end{Abstract}
\end{start}

\section[]{Introduction}

Generalized Parton Distributions (GPDs)\cite{first} 
parametrize the non-perturbative hadron structure
in hard exclusive 
processes.
Their measurement would represent 
a unique way to access several 
crucial features of the nucleon
(for a comprehensive review, 
see, e.g., Ref.\cite{dpr}).
According to a factorization theorem
derived in QCD\cite{fact}, GPDs enter
the long-distance dominated part of
exclusive lepton Deep Inelastic Scattering
(DIS) off hadrons.
In particular, Deeply Virtual Compton Scattering (DVCS),
i.e. the process
$
e H \longrightarrow e' H' \gamma
$ when
$Q^2 \gg m_H^2$,
is one of the the most promising to access GPDs.
As a matter of facts, deeply virtual meson photoproduction
is affected by final state interactions.
Here and in the following,
$Q^2$ is the momentum transfer between the leptons $e$ and $e'$,
and $\Delta^2$ the one between the hadrons $H$ and $H'$
\cite{gui}.
Therefore,
relevant experimental efforts to measure GPDs
by means of DVCS off hadrons are likely to
take place in the next few years.
Recently, the issue of measuring GPDs for nuclei
has been addressed. The first paper on this subject
\cite{cano1}, concerning the deuteron,
contained already the crucial observation that
the knowledge of GPDs would permit the investigation
of the short light-like distance structure of nuclei, and thus the interplay
of nucleon and parton degrees of freedom in the nuclear 
wave function.
In standard DIS off a nucleus
with four-momentum $P_A$ and $A$ nucleons of mass $M$,
this information can be accessed in the 
region where
$A x_{Bj} \simeq {Q^2\over 2 M \nu}>1$,
being $x_{Bj}= Q^2/ ( 2 P_A \cdot q )$ and $\nu$
the energy transfer in the laboratory system.
In this region measurements are difficult, because of 
vanishing cross-sections. As explained in Ref.\cite{cano1}, 
the same physics can be accessed
in DVCS at lower values of $x_{Bj}$.
Since then, DVCS
has been extensively discussed for
nuclear targets.
Calculations
have been performed for
the deuteron\cite{cano2} and for finite nuclei
\cite{gust,kir,liu}. 
The study of GPDs for $^3$He is interesting
for many aspects. 
In fact, $^3$He is a well known nucleus, for which realistic studies 
are possible, so that conventional nuclear effects
can be safely calculated. Strong deviations from the predicted
behaviour could be ascribed to exotic effects, such as
the ones of non-nucleonic degrees of freedom, not included in a
realistic wave function.
Besides, $^3$He is extensively used as an effective neutron target,
in  DIS, in particular in the polarized case \cite{friar,io2}.
Polarized $^3$He will be the first candidate
for experiments aimed at the study of
GPDs of the free neutron, to unveil details of its angular momentum
content. 
In this talk, 
the results of an impulse approximation (IA)
calculation\cite{prc} of the quark unpolarized GPD $H_q^3$ of
$^3$He are reviewed. A convolution formula
is discussed and numerically evaluated
using a realistic non-diagonal spectral function,
so that Fermi motion and binding effects are rigorously estimated.
The proposed scheme is valid for $\Delta^2 \ll Q^2,M^2$
and despite of this it permits to
calculate GPDs in the kinematical range relevant to
the coherent, no break-up channel of deep exclusive processes off $^3$He.
In fact, the latter channel is the most interesting one for its 
theoretical implications, but it can be hardly observed at
large $\Delta^2$, due to the vanishing cross section.
The main result of this investigation
is not the size and shape of the obtained $H_q^3$ for $^3$He,
but the size and nature of nuclear effects on it.
This will permit to test directly, for the
$^3$He  target at least, the accuracy of prescriptions
which have been proposed to estimate nuclear GPDs\cite{kir},
providing a tool for the planning of future experiments
and for their correct interpretation.

\section{Formalism}

The formalism introduced in Ref.\cite{jig} is adopted. 
If one thinks to a spin $1/2$ hadron target, with initial (final)
momentum and helicity $P(P')$ and $s(s')$, 
respectively, two 
GPDs $H_q(x,\xi,\Delta^2)$ and
$E_q(x,\xi,\Delta^2)$, occur.
If one works in a system of coordinates where
the photon 4-momentum, $q^\mu=(q_0,\vec q)$, and $\bar P=(P+P')/2$ 
are collinear along $z$,
$\xi$ is the so called ``skewedness'', parametrizing
the asymmetry of the process, is defined
by the relation 
\bq
\xi = - {n \cdot \Delta \over 2} = - {\Delta^+ \over 2 \bar P^+}
= { x_{Bj} \over 2 - x_{Bj} } + {{O}} \left ( {\Delta^2 \over Q^2}
\right ) ~,
\label{xidef}
\eq
where $n$
is a light-like 4-vector
satisfying the condition $n \cdot \bar P = 1$.
One should notice that the variable $\xi$
is completely fixed by the external lepton kinematics.
The values of $\xi$ which are possible for a given value of
$\Delta^2$ are
$
0 \le \xi \le \sqrt{- \Delta^2}/\sqrt{4 M^2-\Delta^2}~.
$
The well known natural constraints of $H_q(x,\xi,\Delta^2)$ are: 
i) the so called
``forward'' limit, 
$P^\prime=P$, i.e., $\Delta^2=\xi=0$, where one 
recovers the usual PDFs
$
H_q(x,0,0)=q(x)~;
\label{i)}
$
ii)
the integration over $x$, yielding the contribution
of the quark of flavour $q$ to the Dirac 
form factor (f.f.) of the target:
$
\int dx H_q(x,\xi,\Delta^2) = F_1^q(\Delta^2)~;
\label{ii)}
$
iii) the polynomiality property\cite{jig}.

In Ref.\cite{prc}, specifying to the $^3$He target
the procedure developed in
Ref.\cite{io3},
an IA expression
for $H_q(x,\xi,\Delta^2)$ of a given hadron target, 
for small values of $\xi^2$, has been obtained:

\begin{eqnarray}
\label{flux}
H_q^3(x,\xi,\Delta^2) 
& =  & 
\sum_N \int dE \int d \vec p
\, 
[ P_{N}^3(\vec p, \vec p + \vec \Delta, E ) + 
{{O}} 
( {\vec p^2 / M^2},{\vec \Delta^2 / M^2}) ]
\nonumber
\\
& \times & 
{\xi' \over \xi}
H_{q}^N(x',\xi',\Delta^2) + 
{{O}} 
\left ( \xi^2 \right )~.
\label{spec}
\end{eqnarray}
In the above equation, the kinetic energies of the residual nuclear
system and of the recoiling nucleus have been neglected, and
$P_{N}^3 (\vec p, \vec p + \vec \Delta, E )$ is
the one-body off-diagonal spectral function
for the nucleon $N$ in $^3$He:
\begin{eqnarray}
P_N^3(\vec p, \vec p + \vec \Delta, E)  & = & 
{1 \over (2 \pi)^3} {1 \over 2} \sum_M 
\sum_{R,s}
\bra \vec P'M | (\vec P - \vec p) S_R, (\vec p + \vec \Delta) s\ket 
\nonumber
\\
& \times &  
\bra (\vec P - \vec p) S_R,  \vec p s| \vec P M \ket
\, \delta(E - E_{min} - E^*_R)~.
\label{spectral}
\end{eqnarray}
Besides, the quantity
$
H_q^N(x',\xi',\Delta^2)
$
is the GPD of the bound nucleon N
up to terms of order $O(\xi^2)$, and in the above equation
use has been made of
the relations
$
\xi'  =  - \Delta^+ / 2 \bar p^+~,
$
and $ x' = (\xi' / \xi) x$~.

The delta function in Eq. (\ref{spectral})
defines $E$, the so called removal energy, in terms of
$E_{min}=| E_{^3He}| - | E_{^2H}| = 5.5$ MeV and
$E^*_R$, the excitation energy 
of the two-body recoiling system.
The main quantity appearing in the definition
Eq. (\ref{spectral}) is
the overlap integral
\bq
\bra \vec P M | \vec P_R S_R, \vec p s \ket=
\int d \vec y \, e^{i \vec p \cdot \vec y}
\bra \chi^{s},
\Psi_R^{S_R}(\vec x) | \Psi_3^M(\vec x, \vec y) \ket~,
\label{trueover}
\eq 
between the eigenfunction 
$\Psi_3^M$ 
of the ground state
of $^3$He, with eigenvalue $E_{^3He}$ and third component of
the total angular momentum $M$, and the
eigenfunction $\Psi_R^{S_R}$, with eigenvalue
$E_R = E_2+E_R^*$ of the state $R$ of the intrinsic
Hamiltonian pertaining to the system of two interacting
nucleons\cite{over}.
Since the set of the states $R$ also includes
continuum states of the recoiling system, the summation
over $R$ involves the deuteron channel and the integral
over the continuum states.
Eq. (\ref{spec}) can be written in the form
\begin{eqnarray}
H_{q}^3(x,\xi,\Delta^2) =  
\sum_N \int_x^1 { dz \over z}
h_N^3(z, \xi ,\Delta^2 ) 
H_q^N \left( {x \over z},
{\xi \over z},\Delta^2 \right)~,
\label{main}
\end{eqnarray}
where 
\begin{equation}
h_N^3(z, \xi ,\Delta^2 ) =  
\int d E
\int d \vec p
\, P_N^3(\vec p, \vec p + \vec \Delta) 
\delta \left( z + \xi  - { p^+ \over \bar P^+ } \right)~.
\label{hq0}
\end{equation}

In Ref.\cite{prc}, it is discussed that
Eqs. (\ref{main}) and (\ref{hq0}) or, which is the same,
Eq. (\ref{spec}), fulfill the constraint $i)-iii)$ previously listed.

The constraint $i)$, i.e. the forward limit
of GPDs, is certainly verified.
In fact, by taking
the forward limit ($\Delta^2 \rightarrow 0, \xi \rightarrow 0$)
of Eq. (\ref{main}), one gets the 
expression which is usually found,
for the parton distribution $q_3(x)$, in the IA analysis of
unpolarized DIS off $^3He$: 
\begin{eqnarray}
q_3(x) =  H_q^3(x,0,0) =
\sum_{N} \int_x^1 { dz \over z}
f_{N}^3(z) \,
q_{N}\left( {x \over z}\right)~.
\label{mainf}
\end{eqnarray}
In the latter equation,
\begin{equation}
f_{N}^3(z) = h_{N}^3(z, 0 ,0) =  \int d E \int d \vec p
\, P_{N}^3(\vec p,E) 
\delta\left( z - { p^+ \over \bar P^+ } \right)
\label{hq0f}
\end{equation}
is the light-cone momentum distribution of the nucleon $N$
in the nucleus, $q_N(x)= H_q^N( x , 0, 0)$
is the distribution
of the quark of flavour $q$ 
in the nucleon $N$ and $P_N^3(\vec p, E)$,
the $\Delta^2 \longrightarrow 0$ limit of
Eq. (3), is the
one body spectral function.

The constraint $ii)$, i.e. the $x-$integral of the GPD
$H_q$, is also naturally
fulfilled. In fact, by $x-$integrating Eq. (\ref{main}),
one easily obtains:
\begin{eqnarray}
\int dx H_q^3(x,\xi,\Delta^2) & = & \sum_N
\int dx \int {dz \over z} h_N^3(z,\xi,\Delta^2)
H_q^N \left ( { x \over z}, {\xi \over z},
\Delta^2 \right ) =
\nonumber
\\
& = &
\sum_N
\int d x' H_q^N (x',\xi',\Delta^2) \int d z 
h_N^3(z,\xi,\Delta^2) =
\nonumber
\\
& = &
\sum_N 
F_q^N(\Delta^2)
F_N^3(\Delta^2)
= F_q^3(\Delta^2)~.
\label{ffc}
\eq
In the equation above,
$F_q^3(\Delta^2)$ is the
contribution, 
of the quark of flavour $q$,
to the
nuclear f.f.;
$F_q^N(\Delta^2)$ is the contribution,
of the quark of flavour $q$,  
to the nucleon $N$ f.f.;
$F_N^3(\Delta^2)$ is the 
so-called $^3He$ ``pointlike f.f.'', which
would represent the contribution of the nucleon $N$ to the
f.f. of $^3He$ if $N$ were point-like.
$F_N^3(\Delta^2)$ is given, in the present approximation, by
\bq
F_N^3(\Delta^2) = \int dE \int d \vec p
\, P_N^3(\vec p, \vec p + \vec \Delta, E)
= \int dz \, h_N^3(z,\xi,\Delta^2)~. 
\label{ffp}
\eq

Eventually the polynomiality, condition $iii)$,
is formally fulfilled by Eq. (\ref{spec}), although
one should always remember that it is a 
result of order ${\cal{O}}(\xi^2)$,
so that high moments cannot be really checked.

\section{Numerical Results}

$H_q^3(x,\xi,\Delta^2)$, Eq. (\ref{spec}), 
has been evaluated in the nuclear Breit Frame.

The non-diagonal spectral function
Eq. (\ref{spectral}), appearing in Eq.
(\ref{spec}),
has been calculated 
along the lines of Ref.\cite{gema},
by means of 
the overlap Eq. (\ref{trueover}), which 
exactly includes
the final state interactions in the two nucleon recoiling system,
the only plane wave being that describing the relative motion
between the knocked-out nucleon and the two-body system
\cite{over}. 
The realistic wave functions $\Psi_3^M$
and $\Psi_R^{S_R}$ in Eq. (\ref{trueover})
have been evaluated
using the 
AV18 interaction\cite{av18} 
and
taking into account
the Coulomb repulsion of protons in $^3$He.
In particular $\Psi_3^M$ has been 
developed along the lines of Ref.\cite{tre}.
The other ingredient in Eq. (\ref{spec}), i.e.
the nucleon GPD $H_q^N$, has been modelled in agreement with
the Double Distribution representation\cite{rad1}.
In this model, whose details are summarized in Ref.\cite{prc},
the $\Delta^2$-dependence of $H_q^N$ is given by
$F_q(\Delta^2)$, i.e. the contribution
of the quark of flavour $q$
to the nucleon form factor.
It has been obtained from
the experimental values of the proton, $F_1^p$, and
of the neutron, $F_1^n$, Dirac form factors. For the
$u$ and $d$ flavours, neglecting the effect of the 
strange quarks, one has
$
F_u (\Delta^2) =  {1 \over 2} (2 F_1^p(\Delta^2) + F_1^n(\Delta^2))~,
$
$
F_d (\Delta^2) =  2 F_1^n(\Delta^2) + F_1^p(\Delta^2)~.
$
The contributions of 
\begin{figure}[ht]
\centerline{\epsfxsize=2.4in\epsfbox{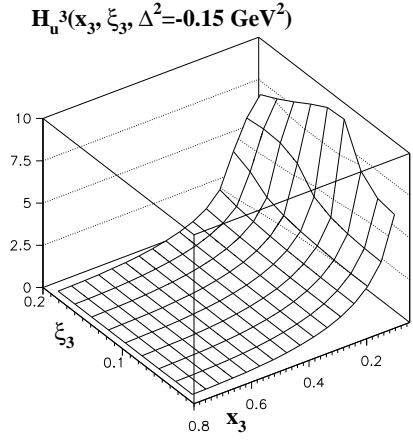}}   
\caption{
For the $\xi_3$ values
which are allowed 
at $\Delta^2 = -0.15$ GeV$^2$,
$H_u^3(x_3,\xi_3,\Delta^2)$,
evaluated using Eq. (\ref{main}),
is shown for $0.05 \leq x_3 \leq 0.8$.}
\end{figure}
the flavours $u$ and $d$
to the proton and neutron f.f. are therefore
$
F_u^p (\Delta^2) =  {4 \over 3} F_u(\Delta^2)~,
$
and
$
F_d^p  =  - {1 \over 3} F_d(\Delta^2) 
$
and
$
F_u^n (\Delta^2) =  
{2 \over 3} F_d(\Delta^2)~,
$
$
F_d^n (\Delta^2)  =  - {2 \over 3} F_u(\Delta^2)~,
$
respectively.
For the numerical calculations,
use has been made of the parametrization of the nucleon
Dirac f.f. given in Ref.\cite{gari}.
Now
the ingredients of the calculation 
have been completely described, so that numerical results
can be presented.
If one considers
the forward limit of the ratio
\bq
R_q (x,\xi,\Delta^2) = 
{ H_q^3(x,\xi,\Delta^2) 
\over 
2 H_q^p(x,\xi,\Delta^2) + H_q^n(x,\xi,\Delta^2)}~,
\label{rat}
\eq
where the denominator clearly represents
the distribution of the
quarks of flavour $q$ 
in $^3$He if nuclear effects are completely
disregarded, i.e., the interacting quarks are assumed to belong
to free nucleons at rest,
the behaviour which is found, shown in Ref.\cite{prc},
is typically $EMC-$like,
so that, 
in the forward limit, well-known results are recovered.
In Ref.\cite{prc} it is also shown that
the $x$ integral of the nuclear GPD gives a good 
description of ff data of $^3$He, in the relevant kinematical region,
$-\Delta^2 \leq 0.25$ GeV$^2$.
As an illustration,
the result of the evaluation of 
$H_u^3(x,\xi,\Delta^2)$
by means of Eq. (\ref{spec})
is shown in Fig. 1, for 
$\Delta^2 =-0.15$ GeV$^2$
as a function of $x_3=3x$ and $\xi_3=3 \xi$.
The GPDs are shown
for the $\xi_3$ range allowed and in the $x_3 \geq 0$ 
\begin{figure}[ht]
\centerline{\epsfxsize=2.2in\epsfbox{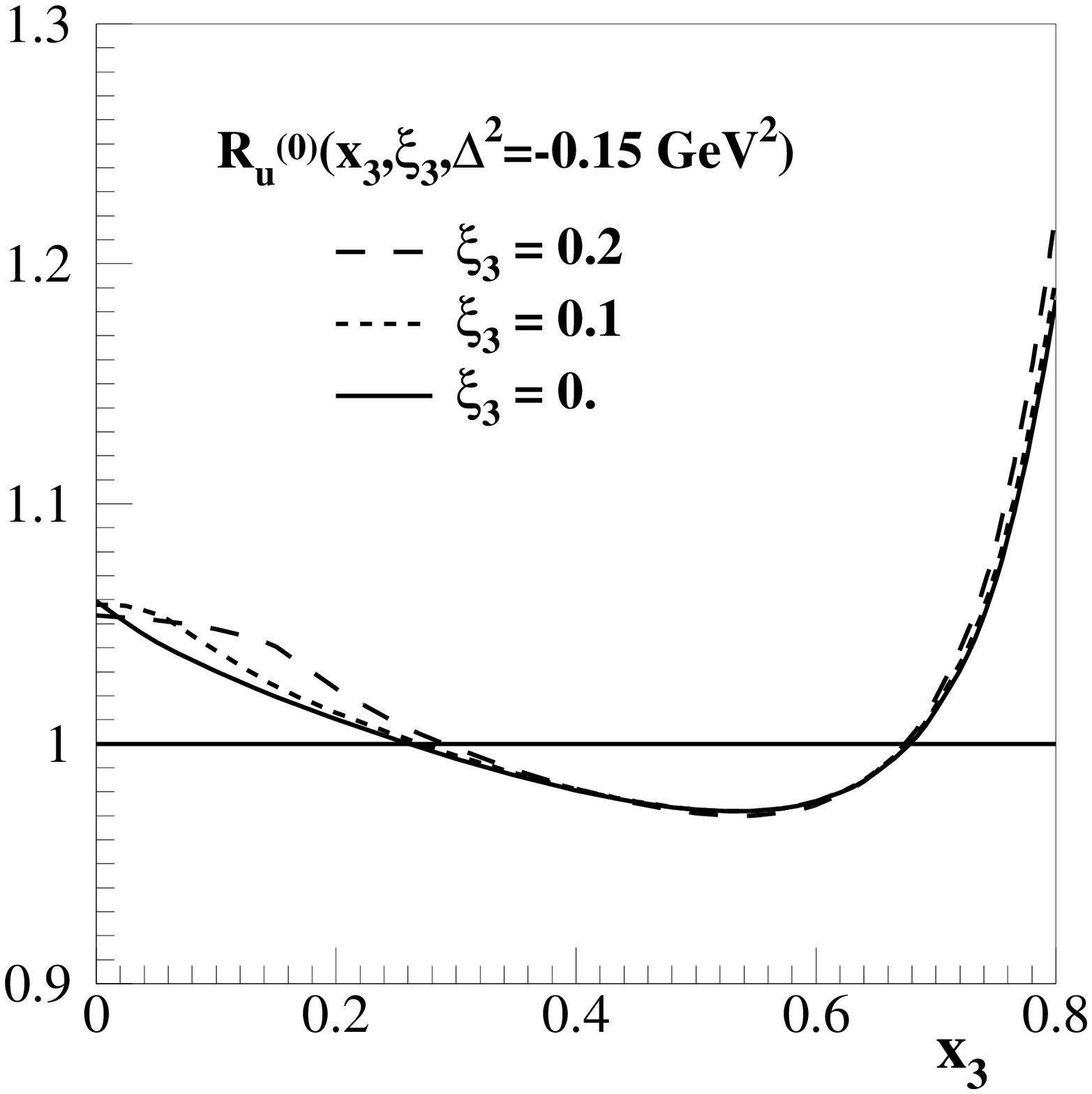}
\epsfxsize=2.2in\epsfbox{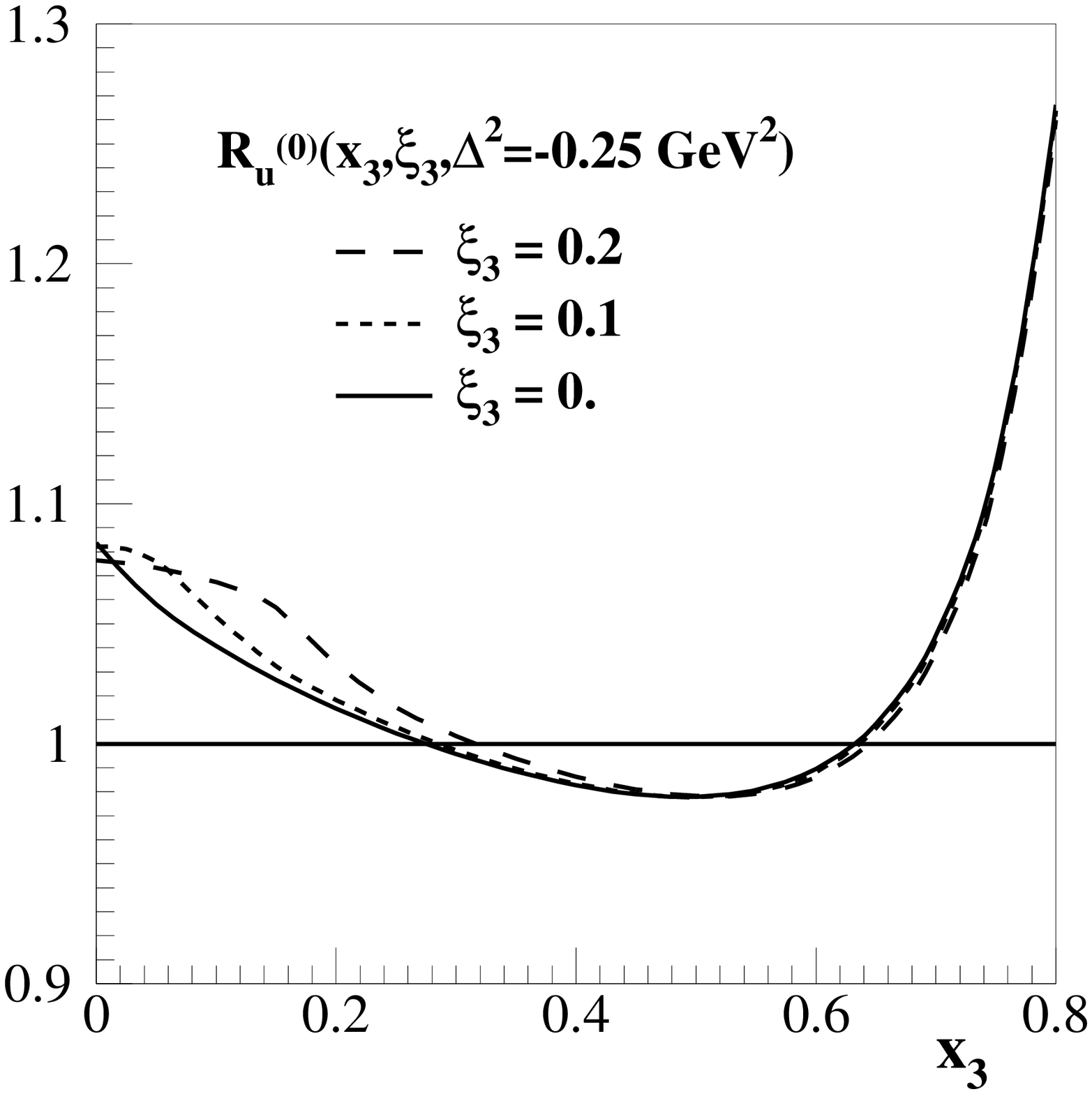}}   
\caption{
In the left panel,
the ratio Eq. (\ref{rnew}) is shown, for the $u$ flavour and
$\Delta^2 = -0.15$ GeV$^2$, 
as a function of $x_3$.
The full line has been calculated for $\xi_3=0$,
the dashed line for $\xi_3=0.1$ and the long-dashed
one for $\xi_3=0.2$. 
The symmetric part at $ x_3 \leq 0$ is not presented.
In the right panel, the same is shown, for  
$\Delta^2 = -0.25$ GeV$^2$. 
}
\end{figure}
region.
Let us now discuss the quality and size of the nuclear effects.
The full result for the GPD $H_q^3$, Eq. (\ref{spec}),
will be now
compared with a prescription
based on the assumptions
that nuclear effects are completely neglected
and the global $\Delta^2$ dependence can be
described 
by
the f.f. of $^3$He:
\bq
H_q^{3,(0)}(x,\xi,\Delta^2) 
= 2 H_q^{3,p}(x,\xi,\Delta^2) + H_q^{3,n}(x,\xi,\Delta^2)~,
\label{app0}
\eq
where the quantity
$
H_q^{3,N}(x,\xi,\Delta^2)=  
\tilde H_q^N(x,\xi)
F_q^3 (\Delta^2)
$
represents the flavor $q$ effective GPD of the bound nucleon 
$N=n,p$ in $^3$He. Its $x$ and $\xi$ dependences, given by the function
$\tilde H_q^N(x,\xi)$, 
is the same of the GPD of the free nucleon $N$,
while its $\Delta^2$ dependence is governed by the
contribution of the quark of flavor $q$ to the
$^3$He f.f., $F_q^3(\Delta^2)$.

The effect of Fermi motion
and binding can be shown through 
the ratio
\be
R_q^{(0)}(x,\xi,\Delta^2) = { H_q^3(x,\xi,\Delta^2) \over H_q^{3,(0)}
(x,\xi,\Delta^2)} 
\label{rnew}
\eq
i.e. the ratio
of the full result, Eq. (\ref{spec}),
to the approximation Eq. (\ref{app0}).
The latter is evaluated by means of the nucleon GPDs used
as input in the calculation, and taking
$
F_u^3(\Delta^2) = {10 \over 3} F_{ch}^{3}(\Delta^2)~,
$
$
F_d^3(\Delta^2) = -{4 \over 3} F_{ch}^{3}(\Delta^2)~,
$
where $F^3_{ch}(\Delta^2)$ is the f.f. which is calculated
within the present approach.
The coefficients $10/3$ and $-4/3$ are simply chosen
assuming that the contribution of the
valence quarks of a given flavour to the f.f. of $^3$He
is proportional to their charge. 
The choice of calculating the ratio Eq. (\ref{rnew})
to show nuclear effects is a very natural one.
As a matter of fact, the forward limit of the ratio Eq. (\ref{rnew})
is the same of the ratio Eq. (\ref{rat}), yielding the
EMC-like ratio for the parton distribution $q$ and,
if $^3$He were made of free nucleons at rest,
the ratio Eq. (\ref{rnew}) would be one.
This latter fact can be immediately realized by
observing that the prescription Eq. (\ref{app0})
is exactly obtained by
placing $z=1$, i.e. no Fermi motion effects
and no convolution, into Eq. (\ref{spec}). 

\begin{figure}[ht]
\centerline{\epsfxsize=2.4in\epsfbox{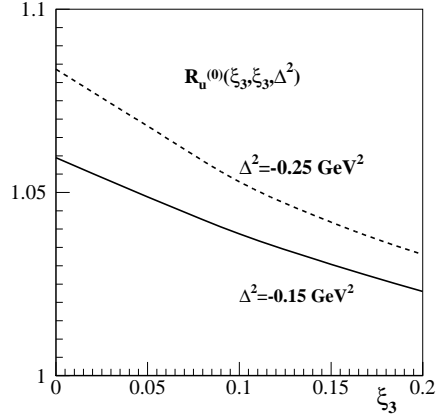}}   
\caption{
The ratio Eq. (\ref{rnew}) 
for the $u$ flavour, for $x_3=\xi_3$,
as a function of $\xi_3$,
at $\Delta^2 = -0.15$ GeV$^2$ (full line),
and at $\Delta^2 = -0.25$ GeV$^2$ (dashed line).
}
\end{figure}

Results are presented in Fig. 2, where
the ratio Eq. (\ref{rnew}) is shown
for $\Delta^2 = -0.15$ GeV$^2$ 
as a function of $x_3$,  
for three different values
of $\xi_3$, for the flavours $u$ and $d$.

Some general trends of the results are apparent:

i) nuclear effects, for $x_3 \leq 0.7$, are as large as 15 \% at most.

ii) Fermi motion and binding have their main effect
for $x_3 \leq 0.3$, at variance with what happens
in the forward limit.

iii) nuclear effects increase with
increasing $\xi$ and
$\Delta^2$, for $x_3 \leq 0.3$.

iv) nuclear effects for the $d$ flavour are larger than
for the $u$ flavour.

The behaviour described above 
can be explained as follows.
As already said in section 2, in IA and in the forward limit, at $x_3=0$
one basically recovers the spectral function normalization
and no nuclear effects, so that the ratio Eq. (\ref{rat}) 
slightly differs from one.
This is not true of course in the present case, due to nuclear
effects hidden not only in the $x'$ dependence, but also
in the $\xi'$ one.
Moreover, even if $x_3=\xi_3=0$, in the present situation
the ratio Eq. (\ref{rnew}) does not give the spectral function 
normalization as in the forward case, 
because of the $\Delta^2$ dependence.
One source of such dependence is that,
in the approximation
Eq. (\ref{app0}), it is assumed that the quarks $u$ and $d$,
belonging to the protons or to the neutron in
$^3He$, contribute to the charge f.f. in the same way,
being the contribution proportional to their charge only.
Actually, the effect of Fermi motion and binding is stronger
for the quarks belonging to the neutron, having the latter
a larger average momentum with respect to the proton \cite{over}. 
This can be seen
noticing that the pointlike f.f.,  Eq. (\ref{ffp}),
for the proton, shows a stronger $\Delta^2$-dependence, with respect to the
neutron one, the difference being 17 \% (23 \%) at $\Delta^2 = - 0.15$
GeV$^2$ ($\Delta^2 = - 0.25$
GeV$^2$). The prescription used in Eq. (12)
could be correct only
if the pointlike f.f. had a similar $\Delta^2$ dependence. 
Besides, nuclear effects studied by means
of the ratio Eq. (\ref{rnew}) at fixed $x$ and $\xi$
depend on $\Delta^2$, showing clearly 
that such a dependence cannot be factorized,
i.e. the nuclear GPD cannot be written as the product
of a $\Delta^2$ dependent and a $\Delta^2$ independent term,
confirming what has been found for the deuteron case
in Ref. \cite{cano2}.
One should notice that,
if factorization were valid, the left and right
panels of Fig. 2 would be equal.
This fact clearly indicates that a model based on the 
assumption of factorization, such as the one of Ref.
\cite{kir}, is not motivated and cannot be used to parametrize
nuclear GPDs for estimates of DVCS cross sections and asymmetries
for light nuclei.
\begin{figure}[ht]
\centerline{\epsfxsize=2.2in\epsfbox{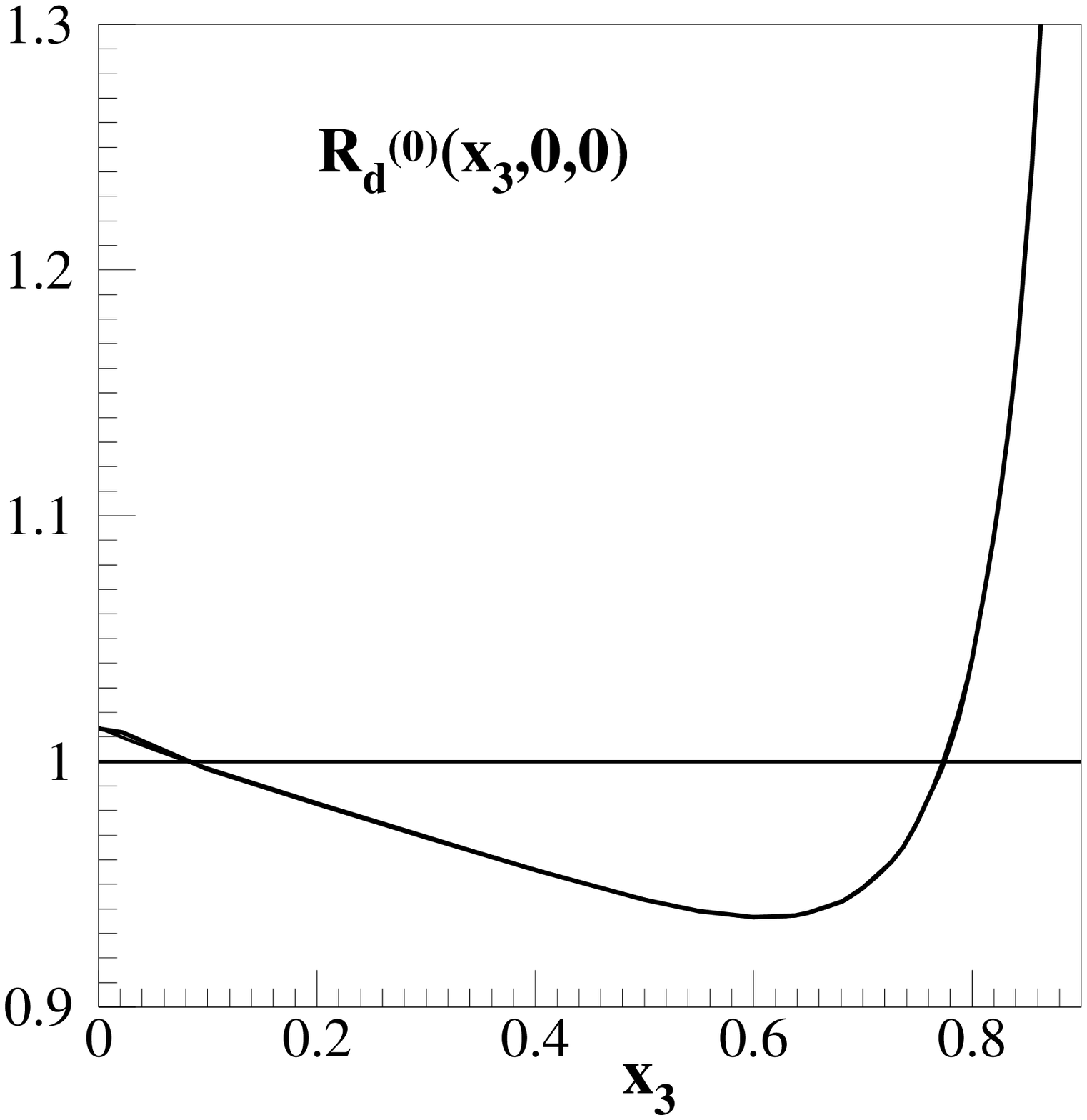}
\epsfxsize=2.2in\epsfbox{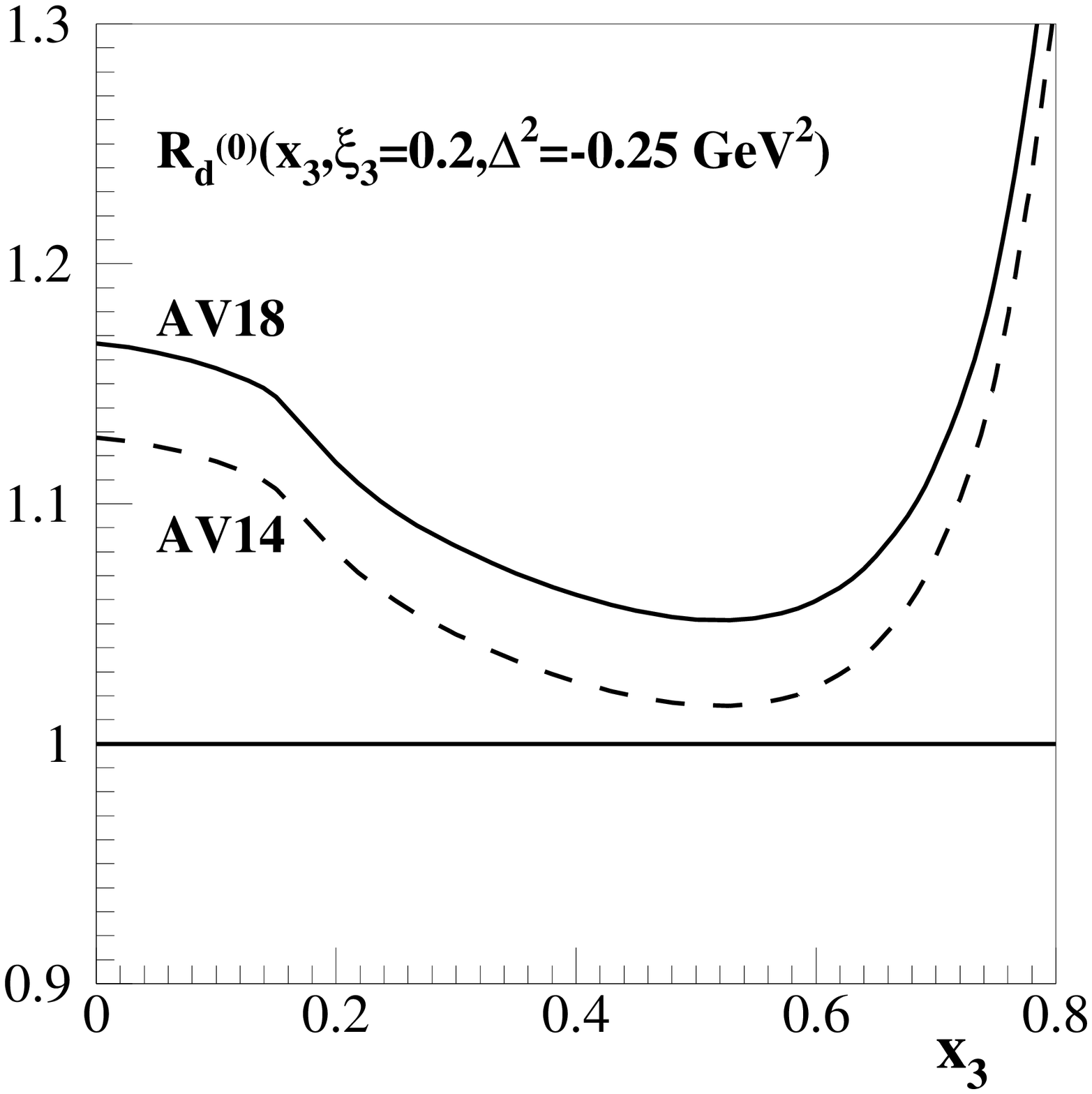}}   
\caption{
Left panel: the ratio $R^{(0)}$, for the
$d$ flavor, in the forward limit $\Delta^2 = 0, \xi=0$, calculated
by means of the AV18 (full line) and AV14
(dashed line) interactions, as 
a function of $x_3 = 3 x$.
The results obtained with the different potentials are
not distinguishable.
Right panel: the same
as in the left panel, but at $\Delta^2=-0.25$ Ge$V^2$ 
and $\xi_3 = 3 \xi = 0.2$. The results are now
clearly distinguishable.
}
\end{figure}
The fact that nuclear effects are larger for the $d$
distribution is also easily explained in terms of
the different contribution of the spectral functions
for the protons and the neutron,
the latter being more important for the GPDs of the $d$ rather than for the
ones of the
$u$ flavour.

A first rough estimate of nuclear effects on 
DVCS observables can be sketched from the obtained results.
In fact, it is known that 
the point $x=\xi$ gives the bulk of the contribution to 
hard exclusive processes,
since at leading order in QCD the amplitude for DVCS and 
for meson electroproduction just involve GPDs at this point.
In Fig.3  it is shown that also in this crucial region
nuclear effects are systematically underestimated
by the approximation Eq. (\ref{app0}).
In Fig. 4, it is shown that nuclear effects 
depend on the choice of the NN potential\cite{gron}, 
at variance with what happens in the forward case.
Nuclear GPDs turn out therefore to be strongly dependent on
the details of nuclear structure.

The issue of applying the obtained GPDs to calculate DVCS off $^3$He, to 
estimate cross-sections and to establish
the feasibility of experiments, 
is in progress. Besides, the study of polarized GPDs
will be very interesting, due to the peculiar
spin structure of $^3$He and its implications
for the study of the angular momentum of the free neutron.

\end{document}